\documentclass[final,5p,times,twocolumn]{elsarticle}

\usepackage{amsfonts,amsmath,amssymb,bm}
\usepackage{graphicx}
\usepackage{color}
\usepackage{textcomp}
\usepackage{subcaption}
\usepackage{ulem}

\definecolor{prdblue}{rgb}{0.133,0.118,0.498}
\usepackage[colorlinks=true, pdfstartview=FitV, linkcolor=prdblue, citecolor= prdblue, urlcolor=prdblue]{hyperref}

\biboptions{sort&compress}
\journal{Physics Letters B}

\graphicspath{
	{./}
	{./figures/}
}

\begin{document}

\begin{frontmatter}

\title{Threshold energies and poles for hadron physical problems by a model-independent universal algorithm}

\author[ECT]{R.-A.~Tripolt}
\author[TEC1]{I.~Haritan}
\author[ECT]{J.~Wambach}
\author[TEC2]{N.~Moiseyev}

\address[ECT]{European Centre for Theoretical Studies in Nuclear Physics and Related Areas (ECT*) and Fondazione Bruno Kessler, Villa Tambosi, Strada delle Tabarelle 286, I-38050 Villazzano (TN), Italy.}
\address[TEC1]{Schulich Faculty of Chemistry, Technion - Israel Institute of Technology, Haifa 32000, Israel}
\address[TEC2]{Schulich Faculty of Chemistry,  Faculty of Physics, and Solid State Institute, Technion - Israel Institute of Technology, Haifa 32000, Israel}

\date{\today}

\begin{abstract}
In this work we show how by using  a Pad\'{e} type analytical continuation scheme, based on the Schlessinger point method, it is possible to find higher production thresholds in hadron physical problems.  Recently, {an} extension of this numerical approach to the complex energy plane enabled the calculations of auto-ionization decay resonance poles in atomic and molecular systems. Here we use this {so-called Resonances via Pad\'{e} (RVP) method}, to show its convergence beyond the singular point in hadron physical problems. In order to demonstrate the capabilities of the RVP method, two illustrations for the ability to identify singularities and branch points are given. In addition, two applications for hadron physical problems are given.  In the first one, we identify the decay thresholds from {a} numerically calculated spectral function. In the second one, we {use experimental data. First, we calculate the resonance pole of the $f_0(500)$ or $\sigma$ meson using the $S0$ partial wave amplitude for $\pi\pi$ scattering in very good agreement with the literature. Second, we use data on the cross section ratio $R(s)$ for $e^+e^-$ collisions and discuss the prediction of decay thresholds which proves to be difficult if the data is noisy.}
\end{abstract}

\end{frontmatter}

\section{Introduction and Motivation}
\label{sec:intro}

The determination of resonance poles, uniquely defined as poles of the $S$-matrix in the complex energy plane, is a long-standing problem and particularly difficult for broad resonances or if decay channels open up in the vicinity. In these cases, simple approaches like a standard Breit-Wigner parametrization fail and more involved theoretical tools like dispersive approaches are necessary, see e.g.~\cite{Oliveothers2014} for reviews. However, these rigorous analytic methods require powerful mathematical techniques which makes them complicated to use in many cases.

In this letter we introduce a method that was originally developed for the calculation of auto-ionization resonances in quantum chemistry \cite{landau2015, landau2016, Idan2016} to the field of hadron physics. This method is model-independent, easy to use and has a broad range of applicability. We refer to this method as the Resonances Via Pad\'{e} (RVP) method. The RVP method is a Pad\'{e} type analytical continuation scheme based on the Schlessinger point method \cite{Schlessinger1966} for calculating resonance poles and threshold energies. The key step in the application of this method is the identification of the analytical domain of the given function. Once this domain is identified, one can use a set of real data points from this domain, and by analytical continuation, calculate resonance poles and predict threshold energies.

Note, that there are different methods to calculate the coefficients in a Pad\'{e} approximate. We use the RVP method based on the Schlessinger point method which is not equivalent to the other Pad\'{e} approximates that are widely used in {a} large variety of fields in physics \cite{MasjuanSanz-Cillero2013,guo_oller,vsvarc2013introducing}.

Let us first explain the common {aspects} between the RVP method, which is based on the Schlessinger point method, and between the Pad\'{e} approximates as used for example in Ref.\cite{MasjuanSanz-Cillero2013}. The input data in the two approaches are values of a function $F(\eta)$ on a {real} grid given by $\{\eta_i\}_{i=0,\pm1 ,\pm2,....}$. The two approaches use the assumption that $\{\eta_i\}_{i=0,\pm1 ,\pm2,....}$ are all located in the analytical domain of the function, to obtain a ratio of two {polynomials}
\begin{equation}
F(\eta)={P(\eta)\over Q(\eta)}.
\label{eq:pade1}
\end{equation}

The main difference between the two methods is  in the range of values of $\eta$ for which the algebraic expansion of $F(\eta)$ is valid.
When the  Pad\'{e} approximates as used for example in Ref.~\cite{MasjuanSanz-Cillero2013}  are used  the expression given in Eq.~\ref{eq:pade1} holds only when $|\eta|< \eta_c$ where $\eta_c$ denotes a singular point of $F(\eta)$ which is closest to the domain of the selected real grid points $\{\eta_i\}_{i=0,\pm1 ,\pm2,....}$. Namely, one can approach the singular point from the ``inside" of the set of the grid points but can not describe $F(\eta)$ beyond $\eta_c$.

However, when the RVP method is used one can describe $F(\eta)$ {also} beyond $\eta_c$ \cite{Schlessinger1966}. Moreover, sufficiently close to $\eta_c$ the expression given in  Eq.~\ref{eq:pade1} obtained by the RVP method shows a non-regular behavior. This ``non-regular" behavior indicates very clearly the region where the singular point $F(\eta_c)$ is located.
This ability is the main message of this paper. It enables us to { study form factors and other observables and look for threshold energies and resonance poles.} Up to our knowledge the convergence of an approximant beyond a singular point is unique to RVP method (see Ref.\cite{Schlessinger1966})   and has not been explored before by other Pad\'{e} approximates.
When given a finite set of $M$ data points $(\eta_i,F_i)$, it is in general not possible to find $F(\eta)$ exactly. We will therefore construct an  approximation to $F(\eta)$ by using the Schlessinger point method \cite{Schlessinger1966}.
The Schlessinger truncated continued fraction $C_M(\eta)$ is then given by
\begin{equation}
C_M(\eta) = \frac{F(\eta_1)}{1+\frac{z_1(\eta-\eta_1)}{1+\frac{z_2(\eta-\eta_2)}{\vdots\, z_{M-1}(\eta-\eta_{M-1})}}},
\label{eq:pade2}
\end{equation}
where the $z_i$ are real coefficients chosen such that
\begin{equation}
C_M(\eta_i)=F(\eta_i),\,\, i= 1,2,\dots, M.
\end{equation}
Once the $z_i$ are determined, an analytic continuation into the complex plane is performed by choosing $\eta$ to be complex, i.e. $\eta=\alpha e^{i\theta}$. For further details on this method and the numerical implementation we refer to \cite{landau2015, Idan2016}.\\

\section{Two illustrations for the ability of the RVP method to identify singularities and branch points}
\label{sec:illustrations}

Let us give a simple example where we compare the two methods. The considered function is
\begin{equation}
F(\eta)=\frac{1}{1-\eta}\,.
\label{F:eq}
\end{equation}
 The input data are {a} set of points within the interval of $0\le \eta < 1$.  The one-pole Pad\'{e} approximant as defined in Eq.~3 in Ref.\cite{MasjuanSanz-Cillero2013} is given by
 \begin{equation}
{\cal{P}}_1^N(\eta,\eta_0=0)=\sum_{n=0}^{N-1}\eta^n+\frac{\eta^N}{1-\eta}\,.
\end{equation}

\begin{figure}[h!]
        \centering
        \begin{subfigure}[h!]{0.98\columnwidth}
	\includegraphics[width=0.99\linewidth]{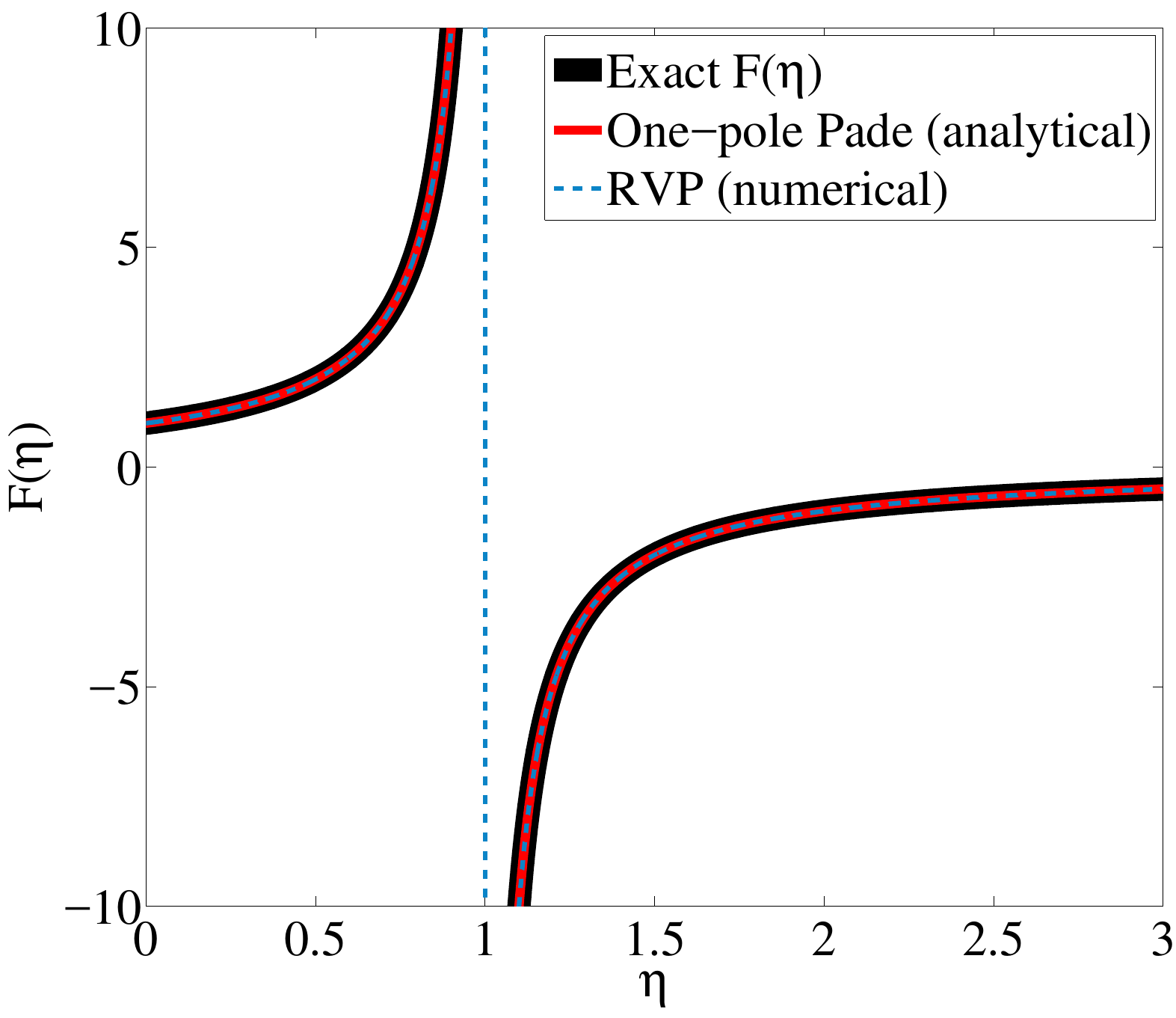}
    \caption{}
	\label{F:analytic}
       \end{subfigure}

     \begin{subfigure}[h!]{0.98\columnwidth}
	\includegraphics[width=0.99\linewidth]{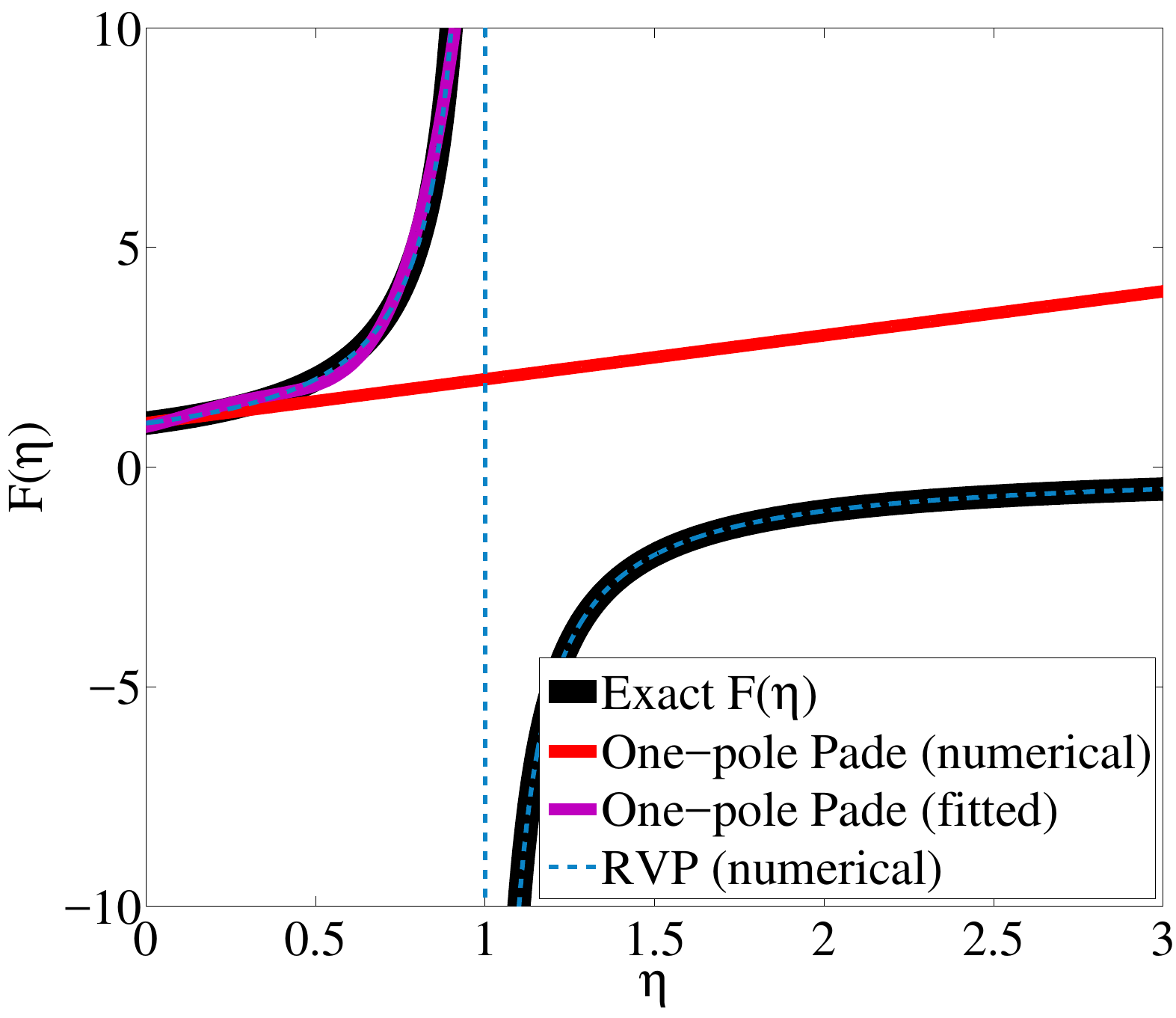}
    \caption{}
	\label{F:numeric}
       \end{subfigure}
       \caption{(color online) Exact and analytically dilated plots for the function $F(\eta)=\frac{1}{1-\eta}$ from Eq.~\ref{F:eq}. (a) Analytical continuation results from the RVP approach (dashed blue line) and from the one-pole Pad\'{e} approach of  Masjuan and Sanz-Cillero with \textbf{analytical} derivatives (red line). Clearly, both methods accurately describe $F(\eta)$ in the whole space, and both accurately describe the singularity. (b) Analytical continuation results from the RVP approach (dashed blue line) and from the one-pole Pad\'{e} approach with \textbf{numerical} derivatives (red line) and with \textbf{fitted} derivatives (purple line). Clearly, both the numerical and fitted one-pole Pad\'{e} approaches fail to discover the singularity and describe $F(\eta)$ after it. Moreover, the numerical one-pole Pad\'{e} approach fails to describe the function even before the singularity.}
       \label{F}
\end{figure}

In Fig.~\ref{F:analytic} we show the results for $N=5$. The excellent agreement with the $F(\eta)$ is expected since ${\cal{P}}_1^{N}(\eta,\eta_0=0)$ is an exact approximation to $F(\eta)$ in {the} whole space for any value of $N$. However, as can be seen from Fig.~\ref{F:numeric}, the one-pole Pad\'{e} approach of Masjuan and Sanz-Cillero, fails to describe $F(\eta)$ close to the singularity region of $F(\eta)$ when the analytical derivatives in Eq.~3 of Ref.~\citenum{MasjuanSanz-Cillero2013} are calculated numerically (around $\eta = 0$, using $dx = 0.0001$) or fitted (using 9 points between 0 to 1, with $R^2=0.9957$).
On the other hand, using the RVP approach the numerical calculations from the same 5-point input data indicate very clearly on the singularity, and describes the correct behavior of $F(\eta)$ far away from the singularity at $\eta = 1$.
This illustrative numerical example shows clearly the advantage of using the RVP numerical approach in the identification of the singularity of an unknown function.

Before studying the application of the RVP approach to hadron physical problems we would like to give another illustrative example to a function of $F(\eta)$ which is non analytical due to a branch point (BP) at $\eta_{BP}=1$:
\begin{equation}
F_{BP}(\eta) = (1-\eta)^{\frac{1}{2}}.
\label{F_BP:eq}
\end{equation}

\begin{figure}[h!]
	\includegraphics[width=0.99\linewidth]{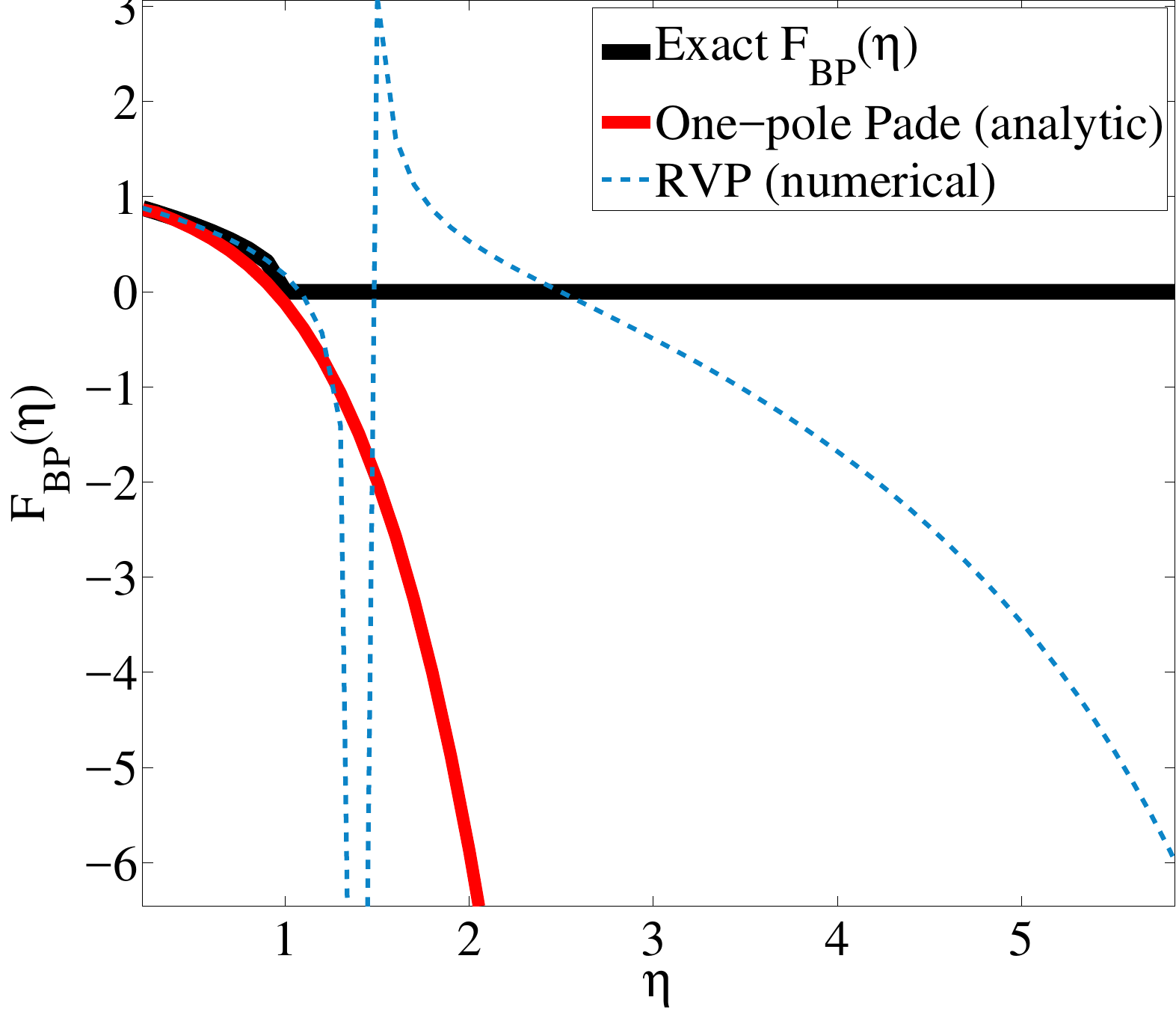}
	\caption{(color online) Exact and analytically dilated plots for the function $Re[F_{BP}(\eta)]=Re[(1-\eta)^{\frac{1}{2}}]$ from Eq.~\ref{F_BP:eq}. Results from the one-pole Pad\'{e} approach of  Masjuan and Sanz-Cillero with \textbf{analytical} derivatives are marked in red, while the results from the numerical RVP approach are marked in dashed blue. Clearly, both methods fail to describe the function after the BP. Yet, while the RVP approach discovers the BP, and exhibits a singular behavior at $\eta\approx1$, the one-pole Pad\'{e} approach fails to discover the BP.}
	\label{F_BP}
\end{figure}

The motivation behind this example is the fact that the BP is often associated with a bifurcation of a particle to two new particles or merging of two particle to a united one particle. In Fig.~\ref{F_BP} we show a comparison between the results obtained by using the one-pole Pad\'{e} approximate with analytical derivatives and between the results obtained by using the numerical RVP approach.
As one can see from Fig.~3, the one-pole Pad\'{e} approximate with analytical derivatives doesn't discover the BP location in spite of the fact that this method works extremely {well} for the discovery of the singularity of $F(\eta)$. However, Fig.~3 clearly shows that the numerical RVP approach discovers quite accurately the BP location even when the input data are 5 grid points which are located far away from the $\eta_{BP}=1$.

\section{Identification  of decay thresholds from  numerical calculated spectral function}
\label{sec:spectral_function}
As a first demonstration of the RVP method we apply it to numerical data on a spectral function in order to identify decay thresholds, i.e. branch points. The underlying theoretical framework to obtain the spectral function are briefly summarized in the following. We wish to emphasize that the focus of this section is to ascertain whether the RVP method can be used as a viable extrapolation method that identifies decay thresholds when using numerical input data.

The spectral function in question has been computed within the quark-meson model in combination with the Functional Renormalization Group (FRG) approach, see \cite{Tripolt2014, Tripolt2014a} for further information. The quark-meson model is a low-energy effective realization of Quantum Chromodynamics (QCD) which involves interactions between quarks, pions and the sigma meson that are compatible with QCD requirements on chiral symmetry and its breaking patterns. We study this model in a thermal medium within the FRG, where the central object is the resolution-scale dependent effective average action, $\Gamma_k$, where $k$ is the renormalization group scale. Its $k=0$ limit, which yields the grand potential, is numerically evolved via a flow equation from the classical action at a large scale $\Lambda$, thereby including quantum and thermal fluctuations. In a similar fashion the (inverse) single-particle propagators, which determine the spectral properties, can be obtained by solving flow equations for the second derivative of the effective action, $\Gamma_k^{(2)}$. The spectral function is then given as
\begin{equation}
\rho(\omega)=-\frac{1}{\pi}\text{Im}\frac{1}{\Gamma^{(2)}_{k\to 0}(\omega)}.
\end{equation}

\begin{figure}[h!]
	\includegraphics[width=0.99\linewidth]{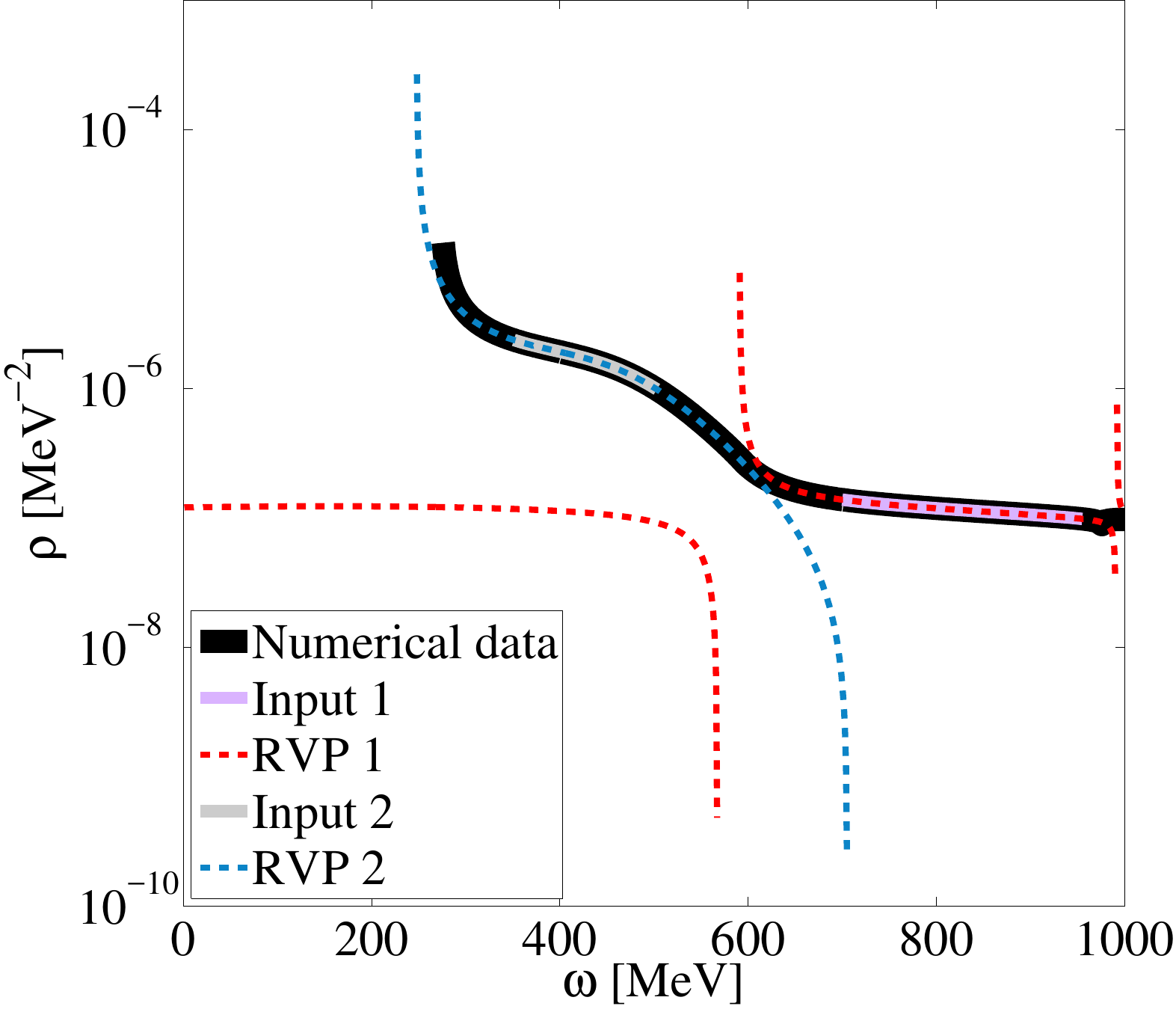}
	\caption{(color online) Numerical and analytically dilated plots for the spectral function of the sigma meson in a thermal medium at temperature $T=60$~MeV. The two dilated graphs (blue and red) are generated through the RVP method, based on different input data (marked in gray and light purple respectively). Clearly, if the input data for the RVP method lies inside the analytic area of the function, it can predict the adjacent BPs. As such, when the input was taken from energies ranging between 700 MeV and 900 MeV (Input1), the RVP method exhibits a singular behavior around 590 MeV and 990 MeV, indicating the approximated location of the BPs. However, when the input was taken from energies ranging between 350 MeV and 500 MeV (Input2), the RVP method exhibited a singular behavior around 250 MeV and 700 MeV, indicating, again, the approximated location of the BPs.}
	\label{fig:model1}
\end{figure}

In the following we focus on results for the spectral function of the sigma meson in a thermal medium at temperature $T$. Fig.~\ref{fig:model1} shows the results for $T=60$~MeV. If the energy of the (off-shell) sigma meson is large enough, it can decay into other particles. When a decay channel opens up or closes, this produces a branch point on the real axis. If the energy is larger than approximately $275$~MeV the sigma meson can decay into two pions, where each have a mass of about $137$~MeV. If the energy is larger than $600$~MeV, it can also decay into a (constituent) quark and an antiquark each having a mass of $300$~MeV. For energies larger than $980$~MeV, a decay into two (on-shell) sigma mesons with a mass of $490$~MeV becomes possible

In summary, we have three decay channels which give rise to three branch points (energy thresholds) on the real axis of the spectral function, namely at 275~MeV, 600~MeV and 980~MeV. The spectral function vanishes below $275$~MeV since there are no decay channels available. The results presented in Fig.~\ref{fig:model1} clearly show how the three high energy thresholds are discovered by the RVP method when the input data are taken far away for these energy thresholds.

\section{Complex pole of the $f_0(500)$- or $\sigma$ meson from experimental data}
\label{sec:sigma}

The identification of scalar mesons and their resonance poles is a long-standing puzzle and particularly difficult for the $f_0(500)$ or $\sigma$ meson due to its large decay width and the strong overlap with the background and higher resonances. For a review on the history and the current status of the $\sigma$ meson we refer to \cite{Pelaez2015}.

In the following we will apply the RVP method  the S0 partial-wave amplitude as obtained from the Constrained Fit to Data (CFD) parametrization of the $\delta_0^{(0)}(s)$ phase shift provided in \cite{Garcia-MartinKaminskiPelaezEtAl2011} which is based on experimental data on $K_{\ell4}$ decays \cite{Pislakothers2001}, in particular the final data from NA48-2 \cite{Batleyothers2010}, and a selection of existing $\pi\pi$ scattering data (see \cite{Garcia-MartinKaminskiPelaezEtAl2011} for details).

Following \cite{Garcia-MartinKaminskiPelaezEtAl2011,Garcia-MartinKaminskiPelaezEtAl2011a}, the partial-wave amplitude for $\pi\pi$ scattering in the $IJ=00$ channel is given by
\begin{align}
t_0^0(s)=\frac{\eta_0^0(s)e^{2i\delta_0^0(s)}-1}{2i\rho_\pi(s)},
\label{eq:t00}
\end{align}
with the phase space factor
\begin{align}
\rho_\pi(s)=\sqrt{1-4M_\pi^2/s}
\end{align}
and the inelasticity $\eta_0^0(s)=1$ for the energy range considered here. The CFD parametrization for the phase shift $\delta_0^{(0)}(s)$ reads
\begin{align}
\cot\delta_0^{(0)}(s)&=\frac{s^{1/2}}{2k}\frac{M_\pi^2}{s-\frac{1}{2}z_0^2}\times\\
&\Bigg(
\frac{z_0^2}{M_\pi\sqrt{s}}+B_0+B_1W(s)+B_2W(s)^2+B_3W(s)^3
\Bigg)\nonumber
\end{align}
with
\begin{align}
W(s)=\frac{\sqrt{s}-\sqrt{s_0-s}}{\sqrt{s}+\sqrt{s_0-s}}, \qquad s_0=4M_K^2,
\end{align}
and
\begin{align}
k(s)=\sqrt{s/4-M_\pi^2}.
\end{align}
The parameters used in these expressions are summarized in Tab.~\ref{tab:S0parameters}.

\begin{table}[h!]
	\begin{center}
		\begin{tabular}{cccc}
			\hline\hline
			$B_0$ & $B_1$ & $B_2$ & $B_3$ \\
			\hline
			$7.14\pm0.23$ &$-25.3\pm0.5$ &$-33.2\pm1.2$ &$-26.2\pm2.3$\\
			\hline
		\end{tabular}
	\end{center}
	\caption{Parameters for the CFD parameterization of the S0 wave phase shift data from
		\cite{Garcia-MartinKaminskiPelaezEtAl2011}. In addition, the pion mass is set to $M_\pi=139.57$~MeV, the kaon mass to $M_K=496$~MeV and $z_0=M_\pi$. We note that this parameterization is only valid up to $\sqrt{s}=850$~MeV.}
	\label{tab:S0parameters}
\end{table}

\begin{figure}[b!]
	\includegraphics[width=\columnwidth]{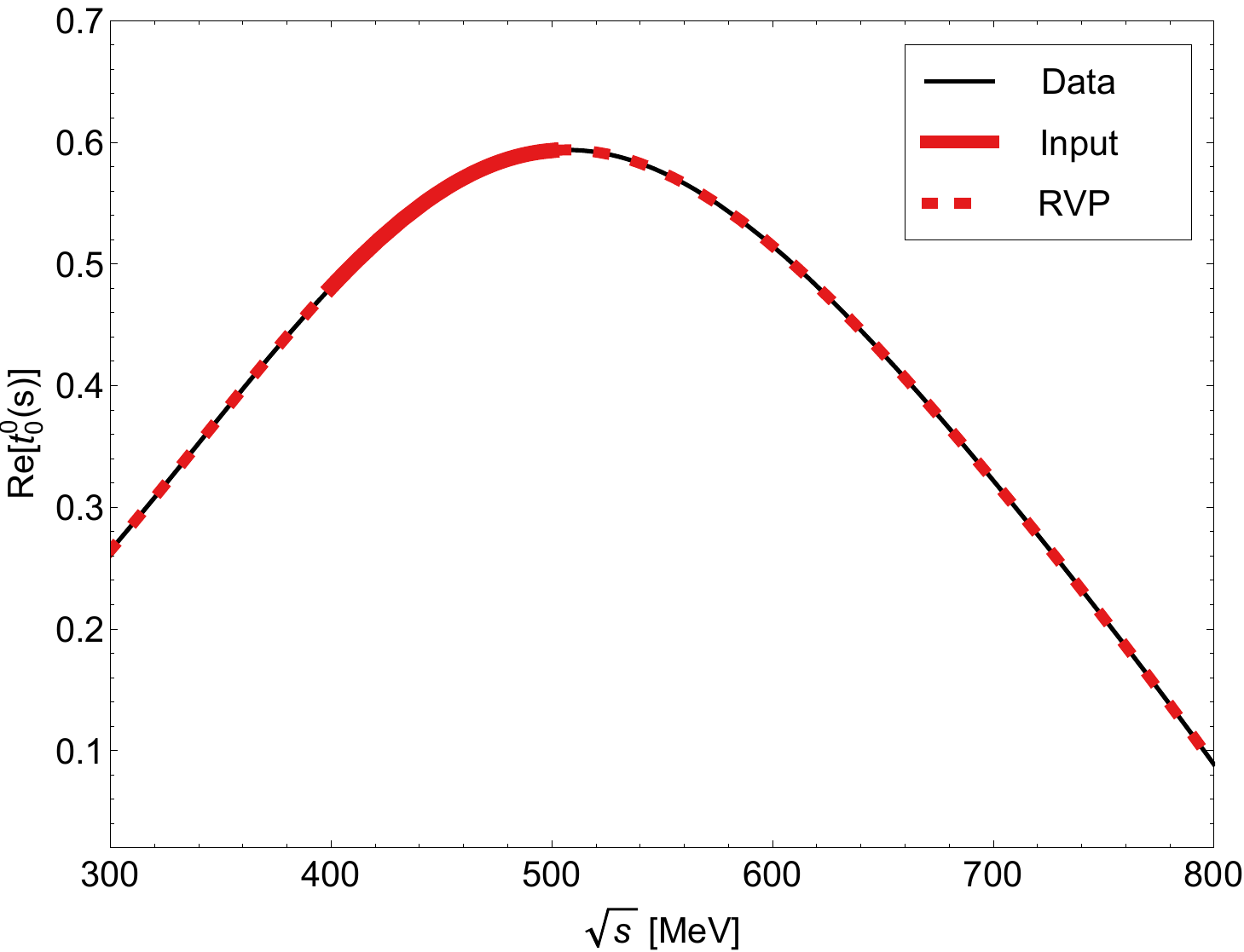}
	\caption{(color online) The real part of the S0 partial wave amplitude, Re$t_0^0(s)$, as obtained from the Constrained Fit to Data (CFD) parametrization of the $\delta_0^{(0)}(s)$ phase shift provided in \cite{Garcia-MartinKaminskiPelaezEtAl2011} is shown together with the chosen input range for the RVP method as well as the obtained extrapolation.}
	\label{fig:Re_t00}
\end{figure}

We will now apply the RVP method to the real part of the partial wave amplitude $t_0^0(s)$ as defined in Eq.~(\ref{eq:t00}). In Fig.~\ref{fig:Re_t00} the real part of $t_0^0(s)$ is shown together with the input region used and the corresponding extrapolation function. We note that it is also possible to use other input regions to determine the complex pole of the $\sigma$ meson since all input regions are generated by the same analytic function which can therefore be reconstructed from any region. In the following we will choose input points from a region between $\sqrt{s}=400$ and $500$~MeV which is closest to the resonance pole in the complex energy plane.

\begin{figure}[h!]
	\includegraphics[width=\columnwidth]{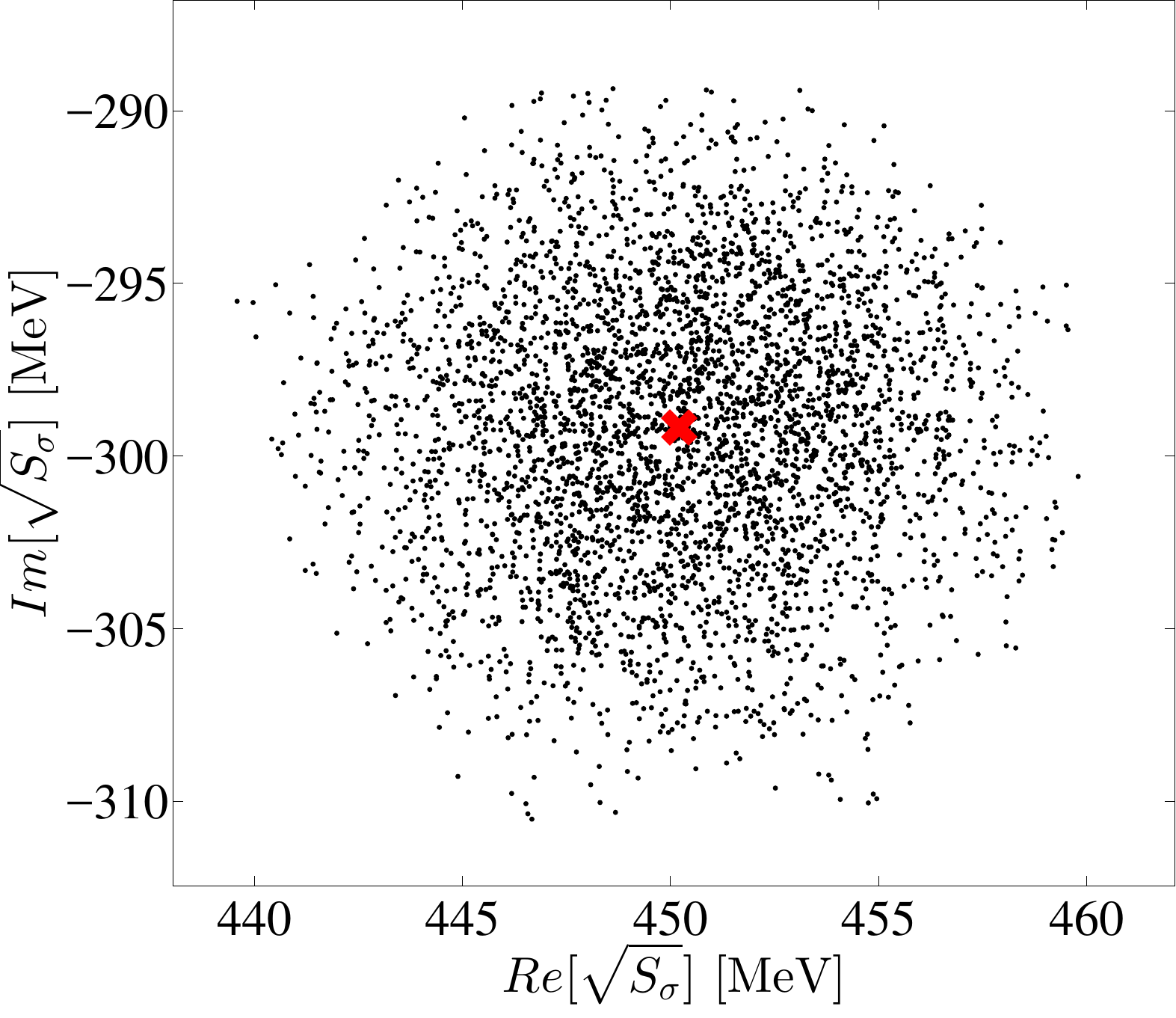}
	\caption{(color online) The calculated real and imaginary values of the $\sigma$ meson complex pole for 3,950 different combinations of the $B_i$ parameters in Tab.~\ref{tab:S0parameters} (black dots). In every combination, each $B_i$ parameter used for the calculation was randomly selected from 1,000 different values ranging from the low uncertainty limit of the $B_i$ to the high uncertainty limit of the $B_i$. The mean value of the $\sigma$ meson complex pole, marked in a red x, was 450.2 -299.2i\,\text{MeV}.}
	\label{fig:errors}
\end{figure}

We find the complex pole of the $\sigma$ meson to be located at
\begin{equation}
\sqrt{s_\sigma}=450.2^{+9.6}_{-10.6} -i(299.2^{+9.8}_{-11.3})\,\text{MeV} .
\end{equation}
To calculate the errors, we created a 1,000 points array for each $B_i$ parameter in Tab.~\ref{tab:S0parameters}. Each array ranged between the low uncertainty value of each $B_i$ to the high uncertainty value of each $B_i$. Later we determined the complex pole for 3,950 random combinations of these $B_i$ parameters (see Fig.~\ref{fig:errors}) and calculated the mean value of the pole and the uncertainty range.

When compared to other predictions for the $\sigma$ resonance pole we find excellent agreement, see Tab.~\ref{tab:sigma_poles}. { Our result should be compared primarily with that of \cite{MasjuanRuizdeElviraSanz-Cillero2014} since it uses the same parameterization for the scattering amplitude as input. All other values quoted in Tab.~\ref{tab:sigma_poles} are based on Roy-type equations with \cite{Garcia-MartinKaminskiPelaezEtAl2011a} representing one of the most recent and advanced dispersive determinations. }

\begin{table}[h]
	\begin{center}
		\begin{tabular}{cccc}
			\hline\hline
			$\sqrt{s_\sigma}$~(MeV)&\hspace{.85cm}source\hspace{.85cm} \\
		\hline
			$470\pm 30-i(295\pm 20)$ & \cite{ColangeloGasserLeutwyler2001} \\[0.5mm]
		%	$470\pm 50-i(285\pm 25)$ & \cite{ZhouQinZhangEtAl2005}  \\[0.5mm]
			$441^{+16}_{-8}-i(272^{+9}_{-12.5})$ & \cite{CapriniColangeloLeutwyler2006}  \\[0.5mm]
			$457^{+14}_{-13}-i(279^{+11}_{-7})$ & \cite{Garcia-MartinKaminskiPelaezEtAl2011a} \\[0.5mm]
			$442^{+5}_{-8}-i(274^{+6}_{-5})$ &\cite{Moussallam2011}\\[0.5mm]
			$449^{+22}_{-16}-i(275\pm 12)$ & \cite{Pelaez2015} \\[0.5mm]
			$453\pm 15-i(297\pm 15)$ & \cite{MasjuanRuizdeElviraSanz-Cillero2014} \\[0.5mm]
			$450.2^{+9.6}_{-10.6} -i(299.2^{+9.8}_{-11.3})$ &this work\\
			\hline
		\end{tabular}
	\end{center}
	\caption{Collection of pole parameter predictions for the $f_0(500)$ or $\sigma$ meson.}
	\label{tab:sigma_poles}
\end{table}

\section{Prediction of decay thresholds for $e^+e^-$ annihilation from experimental data}
\label{sec:ee}

As a final application of the RVP method we will use it to analyze data from $e^+e^-$ collisions and discuss its ability to predict decay (or rather production) thresholds based on experimental data. In particular, we will analyze data on the ratio $R(s)$ between the total cross sections of $e^+e^-$ into hadrons and into muons,
\begin{equation}
R(s)=\frac{\sigma(e^+e^-\rightarrow hadrons)}{\sigma(e^+e^-\rightarrow \mu^+\mu^-)},
\end{equation}
where $\sigma(e^+e^-\rightarrow hadrons)$ is the experimental cross section corrected for initial state radiation and electron-positron vertex loops, and $\sigma(e^+e^-\rightarrow \mu^+\mu^-)=4\pi \alpha^2(s)/3s$ with the electromagnetic fine-structure constant $\alpha(s)$. Depending on the collision energy, different flavors of quarks can be produced.

\begin{figure}[h]
	\includegraphics[width=0.48\textwidth]{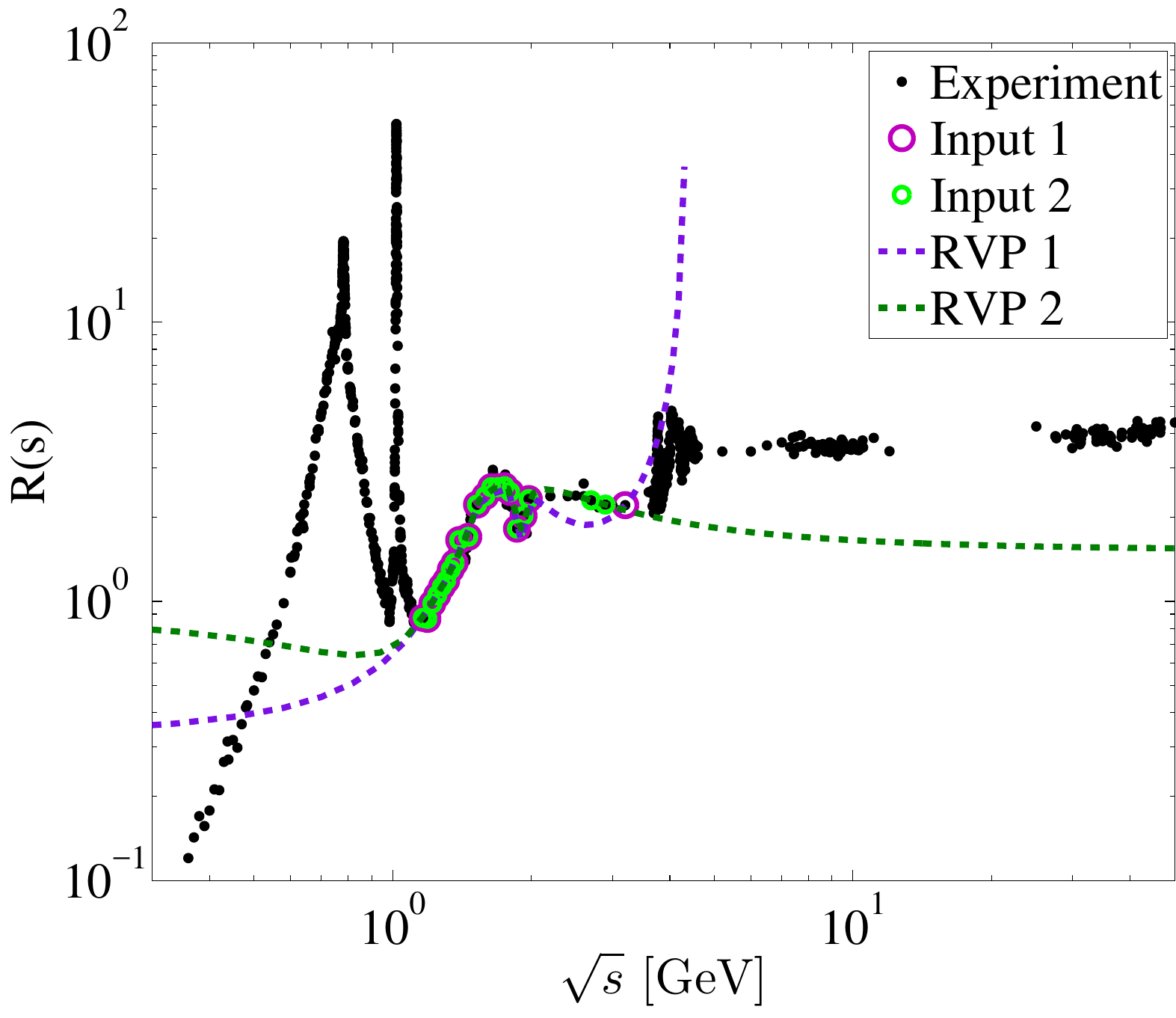}
	\caption{(color online) Collection of data on the ratio $R(s)$ between the total cross sections of $e^+e^-$ into hadrons and into muons, $R(s)=\sigma(e^+e^-\rightarrow hadrons)/\sigma(e^+e^-\rightarrow \mu^+\mu^-)$ from \cite{Oliveothers2014}. Also shown are two input regions chosen for the RVP method together with the resulting extrapolations.}
	\label{fig:ee}
\end{figure}

A collection of data on the ratio $R(s)$ is shown in Fig.~\ref{fig:ee} together with two input regions used for the RVP method and the obtained extrapolation functions. We note that $R(s)$ exhibits a significant increase at $\sqrt{s}\approx 4$~GeV which is related to the production threshold of charm quarks, in particular of $D$ mesons, and the charmonium poles
(see e.g. \cite{ZeninEzhelaLugovskyEtAl2001} for details). As shown in Fig.~\ref{fig:ee}, the two results obtained by the RVP method from the different input data indicate on the singularity at the same energy ($\approx 4$~GeV) by deviation from the given $R(s)$ experimental data. In one extrapolation (marked by green in Fig.~\ref{fig:ee}), the function indicates the threshold energy, although it has a smooth behavior. In the other extrapolation (marked by purple in Fig.~\ref{fig:ee}), the function indicates the threshold energy by exhibiting a singular behavior as in Figs.~\ref{F}-\ref{fig:model1}.

The prediction of thresholds in this case has of course to be treated with care. First of all, the experimental data are noisy which gives rise to a strong dependence on the chosen input points. Moreover, there are several smaller decay thresholds and resonance peaks present in the vicinity of the input regions chosen, which limits the radius of convergence of the obtained Pad\'{e} extrapolations.

We therefore conclude that a robust prediction of decay thresholds is not possible in the present case. We note, however, that the RVP method discussed in this letter is in principle capable of predicting decay thresholds if the input is precise enough and if there is a sufficient number of input points available, as demonstrated for the numerical calculated spectral function in Sec.~\ref{sec:spectral_function}.

\section{Summary}
\label{sec:summary}

In this letter we have introduced a method that was originally developed in \cite{landau2015, landau2016, Idan2016} for the calculation of auto-ionization atomic and molecular resonances in quantum chemistry to hadron physics with the aim to to show the convergence of the RVP numerical approach beyond the singular point,   to predict decay thresholds, and to identify resonance poles. This method, based on the Pad\'{e} approximant, only requires real input in order to reconstruct the underlying function not only along the real axis but also in the complex plane within a certain radius of convergence. The method is universal being model independent, easy to use, and it has a broad range of applicability. We refer to this method as the Resonances Via Pad\'{e} (RVP) method.

In order to demonstrate the abilities of this method we have applied it to several different situations. First, two illustrations for the ability to identify singularities and branch points are given which is hard to find by other Pad\'{e} numerical approaches.  In addition, two applications for hadron physical problems are given.  In the first one, we identify the decay thresholds from numerical calculated spectral function. In the second one, we calculate the energy thresholds and resonance decay poles from experimental data in two cases: the $S0$ partial wave amplitude for $\pi\pi$ scattering and the cross section ratio $R(s)$ for $e^+e^-$ collisions. The extracted values for the resonance poles of the $f_0(500)$ or $\sigma$ meson are in very good agreement with the literature. When the data are noisy the prediction of decay thresholds proves to be less accurate but feasible.

We believe that this method does not only represent a viable tool to improve or supplement current determinations of resonance poles but that it can also be applied to a variety of other situations. In future applications we will further explore its potential and intend to use it, for example, to determine the temperature dependence of resonance poles for hadrons in a hot and dense medium or to obtain real-time correlation functions from their imaginary-time counterparts.

\bigskip

\noindent{{\it Acknowledgements.}} The authors thank A.~Richter for fruitful discussions and for helping to initiate this project. We would also like to thank P.~Masjuan for his clarification of a method developed in \cite{MasjuanRuizdeElviraSanz-Cillero2014} for calculating complex pole parameters. NM and IH acknowledge the I-Core: the Israeli Excellence Center ``Circle of Light", and the Israel Science Foundation grant No. 1530/15 for a partial support of this research. One of us (NM) wishes to express his gratitude to the members of ECT* (where this work has been initiated) for their most warm hospitality during his stay there as a visiting scientist in the summer of 2016.

\bigskip

\noindent\textbf{References}

\smallskip

\bibliographystyle{elsarticle-num}
\bibliography{qcd,pade}

\begin{thebibliography}{10}
\expandafter\ifx\csname url\endcsname\relax
  \def\url#1{\texttt{#1}}\fi
\expandafter\ifx\csname urlprefix\endcsname\relax\def\urlprefix{URL }\fi
\expandafter\ifx\csname href\endcsname\relax
  \def\href#1#2{#2} \def\path#1{#1}\fi

\bibitem{Oliveothers2014}
K.~A. Olive, et~al., {Review of Particle Physics}, Chin. Phys. C38 (2014)
  090001.
\newblock \href {http://dx.doi.org/10.1088/1674-1137/38/9/090001}
  {\path{doi:10.1088/1674-1137/38/9/090001}}.

\bibitem{landau2015}
A.~Landau, I.~Haritan, P.~R. Kapr{\'a}lov{\'a}-Zd{\'a}nsk{\'a}, N.~Moiseyev,
  Atomic and molecular complex resonances from real eigenvalues using standard
  (hermitian) electronic structure calculations, The Journal of Physical
  Chemistry A 120~(19) (2016) 3098--3108.

\bibitem{landau2016}
A.~Landau, D.~Bhattacharya, I.~Haritan, A.~Ben-Asher, N.~Moiseyev, Chapter
  fifteen-ab initio complex potential energy surfaces from standard quantum
  chemistry packages, Advances in Quantum Chemistry 74 (2017) 321--346.

\bibitem{Idan2016}
I.~Haritan?, N.~Moiseyev, On the calculation of resonances by analytic
  continuation of eigenvalues from the stabilization graph, The Journal of
  Chemical Physics 147~(1) (2017) 014101.

\bibitem{Schlessinger1966}
L.~Schlessinger, Use of analyticity in the calculation of nonrelativistic
  scattering amplitudes, Physical Review 167~(5) (1968) 1411--1423.

\bibitem{MasjuanSanz-Cillero2013}
P.~Masjuan, J.~J. Sanz-Cillero, {Pade approximants and resonance poles}, Eur.
  Phys. J. C73 (2013) 2594.
\newblock \href {http://arxiv.org/abs/1306.6308} {\path{arXiv:1306.6308}},
  \href {http://dx.doi.org/10.1140/epjc/s10052-013-2594-4}
  {\path{doi:10.1140/epjc/s10052-013-2594-4}}.

\bibitem{guo_oller}
Z.-H. Guo, J.~Oller, Probabilistic interpretation of compositeness relation for
  resonances, Physical Review D 93~(9) (2016) 096001.

\bibitem{vsvarc2013introducing}
A.~{\v{S}}varc, M.~Had{\v{z}}imehmedovi{\'c}, H.~Osmanovi{\'c}, J.~Stahov,
  L.~Tiator, R.~L. Workman, Introducing the pietarinen expansion method into
  the single-channel pole extraction problem, Physical Review C 88~(3) (2013)
  035206.

\bibitem{Tripolt2014}
R.-A. Tripolt, N.~Strodthoff, L.~von Smekal, J.~Wambach, {Spectral Functions
  for the Quark-Meson Model Phase Diagram from the Functional Renormalization
  Group}, Phys.Rev. D89 (2014) 034010.
\newblock \href {http://arxiv.org/abs/1311.0630} {\path{arXiv:1311.0630}},
  \href {http://dx.doi.org/10.1103/PhysRevD.89.034010}
  {\path{doi:10.1103/PhysRevD.89.034010}}.

\bibitem{Tripolt2014a}
R.-A. Tripolt, L.~von Smekal, J.~Wambach, {Flow equations for spectral
  functions at finite external momenta}, Phys.Rev. D90~(7) (2014) 074031.
\newblock \href {http://arxiv.org/abs/1408.3512} {\path{arXiv:1408.3512}},
  \href {http://dx.doi.org/10.1103/PhysRevD.90.074031}
  {\path{doi:10.1103/PhysRevD.90.074031}}.

\bibitem{Pelaez2015}
J.~R. Pelaez, {From controversy to precision on the sigma meson: a review on
  the status of the non-ordinary $f_0(500)$ resonance}\href
  {http://arxiv.org/abs/1510.00653} {\path{arXiv:1510.00653}}.

\bibitem{Garcia-MartinKaminskiPelaezEtAl2011}
R.~Garcia-Martin, R.~Kaminski, J.~R. Pelaez, J.~Ruiz~de Elvira, F.~J. Yndurain,
  {The Pion-pion scattering amplitude. IV: Improved analysis with once
  subtracted Roy-like equations up to 1100 MeV}, Phys. Rev. D83 (2011) 074004.
\newblock \href {http://arxiv.org/abs/1102.2183} {\path{arXiv:1102.2183}},
  \href {http://dx.doi.org/10.1103/PhysRevD.83.074004}
  {\path{doi:10.1103/PhysRevD.83.074004}}.

\bibitem{Pislakothers2001}
S.~Pislak, et~al., {A New measurement of K+(e4) decay and the s wave pi pi
  scattering length a0(0)}, Phys. Rev. Lett. 87 (2001) 221801, [Erratum: Phys.
  Rev. Lett.105,019901(2010)].
\newblock \href {http://arxiv.org/abs/hep-ex/0106071}
  {\path{arXiv:hep-ex/0106071}}, \href
  {http://dx.doi.org/10.1103/PhysRevLett.105.019901,
  10.1103/PhysRevLett.87.221801} {\path{doi:10.1103/PhysRevLett.105.019901,
  10.1103/PhysRevLett.87.221801}}.

\bibitem{Batleyothers2010}
J.~R. Batley, et~al., {Precise tests of low energy QCD from K(e4)decay
  properties}, Eur. Phys. J. C70 (2010) 635--657.
\newblock \href {http://dx.doi.org/10.1140/epjc/s10052-010-1480-6}
  {\path{doi:10.1140/epjc/s10052-010-1480-6}}.

\bibitem{Garcia-MartinKaminskiPelaezEtAl2011a}
R.~Garcia-Martin, R.~Kaminski, J.~R. Pelaez, J.~Ruiz~de Elvira, {Precise
  determination of the f0(600) and f0(980) pole parameters from a dispersive
  data analysis}, Phys. Rev. Lett. 107 (2011) 072001.
\newblock \href {http://arxiv.org/abs/1107.1635} {\path{arXiv:1107.1635}},
  \href {http://dx.doi.org/10.1103/PhysRevLett.107.072001}
  {\path{doi:10.1103/PhysRevLett.107.072001}}.

\bibitem{MasjuanRuizdeElviraSanz-Cillero2014}
P.~Masjuan, J.~Ruiz~de Elvira, J.~J. Sanz-Cillero, {Precise determination of
  resonance pole parameters through Padé approximants}, Phys. Rev. D90~(9)
  (2014) 097901.
\newblock \href {http://arxiv.org/abs/1410.2397} {\path{arXiv:1410.2397}},
  \href {http://dx.doi.org/10.1103/PhysRevD.90.097901}
  {\path{doi:10.1103/PhysRevD.90.097901}}.

\bibitem{ColangeloGasserLeutwyler2001}
G.~Colangelo, J.~Gasser, H.~Leutwyler, {$\pi \pi$ scattering}, Nucl. Phys. B603
  (2001) 125--179.
\newblock \href {http://arxiv.org/abs/hep-ph/0103088}
  {\path{arXiv:hep-ph/0103088}}, \href
  {http://dx.doi.org/10.1016/S0550-3213(01)00147-X}
  {\path{doi:10.1016/S0550-3213(01)00147-X}}.

\bibitem{CapriniColangeloLeutwyler2006}
I.~Caprini, G.~Colangelo, H.~Leutwyler, {Mass and width of the lowest resonance
  in QCD}, Phys. Rev. Lett. 96 (2006) 132001.
\newblock \href {http://arxiv.org/abs/hep-ph/0512364}
  {\path{arXiv:hep-ph/0512364}}, \href
  {http://dx.doi.org/10.1103/PhysRevLett.96.132001}
  {\path{doi:10.1103/PhysRevLett.96.132001}}.

\bibitem{Moussallam2011}
B.~Moussallam, {Couplings of light I=0 scalar mesons to simple operators in the
  complex plane}, Eur. Phys. J. C71 (2011) 1814.
\newblock \href {http://arxiv.org/abs/1110.6074} {\path{arXiv:1110.6074}},
  \href {http://dx.doi.org/10.1140/epjc/s10052-011-1814-z}
  {\path{doi:10.1140/epjc/s10052-011-1814-z}}.

\bibitem{ZeninEzhelaLugovskyEtAl2001}
O.~V. Zenin, V.~V. Ezhela, S.~B. Lugovsky, M.~R. Whalley, K.~Kang, S.~K. Kang,
  A compilation of total cross-section data on e+ e- into hadrons and pqcd
  tests.\href {http://arxiv.org/abs/hep-ph/0110176}
  {\path{arXiv:hep-ph/0110176}}.

\end{thebibliography}

\end{document}